\documentclass[twocolumn,preprintnumbers,amsmath,amssymb]{revtex4}
\usepackage{epsfig}

\usepackage{graphicx}
\usepackage{dcolumn}
\usepackage{bm}
\usepackage{amsmath}

\newcommand{\be}{\begin{equation}}
\newcommand{\ee}{\end{equation}}

\begin{document}


\title{Is the transition redshift a new cosmological number?}


\author{J. A. S. Lima$^{1}$}\email{limajas@astro.iag.usp.br}
\author{J. F. Jesus$^{1,5}$}\email{jfjesus@itapeva.unesp.br}
\author{R. C. Santos$^{2}$}\email{cliviars@astro.iag.usp.br}
\author{M. S. S. Gill$^{3,4}$}\email{msgill@astronomy.ohio-state.edu}

\vspace{0.5cm}
\affiliation{$^{1}$Departamento de Astronomia, Universidade de S\~ao Paulo, R. do Mat\~ao 1226, 05508-900, S\~ao Paulo, SP, Brazil}
\affiliation{$^{2}$Departamento de Ci\^encias Exatas e da Terra, Universidade Federal de S\~ao Paulo, 09972-270, Diadema, SP, Brazil }

\affiliation{$^{3}$Kavli Institute for Particle Astrophysics \& Cosmology, Stanford, USA}

\affiliation{$^{4}$Center for Cosmology and Astro-Particle Physics, The Ohio State University, \\  
191 West Woodruff Avenue, Columbus, OH 43210, USA}

\affiliation{$^{5}$Universidade Estadual Paulista ``J\'ulio de Mesquita Filho'', Campus Itapeva, R. Geraldo Alckmin 519, 18409-010, Itapeva, SP, Brazil}



\def\zt{\mbox{$z_t$}}

\vspace{0.5cm}
\begin{abstract}

Observations from Supernovae Type Ia (SNe Ia) provided strong evidence
for an expanding accelerating Universe at intermediate redshifts.  This means
that the Universe underwent a transition from
deceleration to acceleration phases at a transition redshift $z_t$ of the
order unity whose value in principle depends on the cosmology as
well as on the assumed gravitational theory. Since cosmological accelerating models
endowed with a transition redshift are extremely degenerated, in principle,  
it is interesting to know whether the value of $z_t$ itself can be observationally used as a
new cosmic discriminator.  After a brief discussion of the potential dynamic
role played by the transition redshift, it is argued that future observations
combining SNe Ia, the line-of-sight (or ``radial'') baryon acoustic oscillations,
the differential age of galaxies, as well as the redshift drift of the spectral
lines may tightly constrain $z_t$, thereby helping to narrow the parameter space
for the most realistic models describing the accelerating Universe.

\vspace{0.1cm}
\end{abstract}

\maketitle

\newpage


\vspace{0.1cm}



\section{Introduction} The extension of the Hubble diagram to larger distances by using
observations from supernovae type Ia (SNe Ia) as standard candles
allowed the history of cosmic expansion to be probed at much higher
accuracy at low and intermediate redshifts.  Independent measurements by
various groups indicated that the current expansion is in fact
speeding up and not slowing down, as was believed for many decades
\cite{sn98,const,Union2,Union21}.  In other words, by virtue of some
unknown mechanism, the expansion of the Universe underwent a ``dynamic
 transition" whose effect was to change the sign of the universal
deceleration parameter $q(z)$.

The correct physical explanation for such a transition is the most
profound challenge for cosmology today. Within the General
Relativistic (GR) paradigm, the simplest manner for explaining such a
phenomenon is by postulating a cosmological constant $\Lambda$ in the
Einstein equations. Indeed, anything which contributes to a decoupled
vacuum energy density also behaves like a cosmological constant.
However, the existence of the so-called cosmological constant and
coincidence problems \cite{weinb}, inspired many
authors to consider alternative candidates, thereby postulating the
existence of an exotic fluid with negative pressure (in addition to
cold dark matter), usually called dark energy
\cite{rev1}.

Possible theoretical explanations for the present
accelerating stage without dark energy are also surprisingly abundant \cite{SF2010,Accel}.
Even in the framework of GR there are some alternative proposals where
the existence of a new dark component is not necessary in order to
have an accelerating regime at low redshifts. For instance, many
authors have claimed that the fact that the acceleration comes out
very close to the beginning of the nonlinear evolution of the contrast
density is not just a trivial coincidence \cite{TB03}.  In
this connection, several averaging procedures have been developed in
order to take into account a possible ``back reaction" effect
associated with the existence of inhomogeneities
\cite{inhom1}. Another possibility still
within the GR framework is that `dynamic transition' can be powered
uniquely by the gravitationally-induced creation of cold dark matter
particles \cite{lima96,BL}. The basic idea is that the
irreversible process of cosmological particle creation at the expense
of the gravitational field can phenomenologically be described by a
negative pressure and the associated entropy production
\cite{prigogine}. Another possibility is provided by models with interaction in the
dark sector, as happens, for instance, in
decaying vacuum cosmologies \cite{decaying1}, as well as, in many
variants of coupled dark energy models \cite{coupled}.

In this paper we advocate a different approach based on the simple existence of a transition
redshift ($z_t$) as required by the SNe Ia data. Its leitmotiv is summarized in the caption of Figure 1.
In our view, due to the recent advances of astronomical
observations such a quantity defining the transition between a decelerating to an accelerating stage will become a powerful cosmological probe.
In particular, it is  argued that future observations combining SNe Ia, the line-of-sight (or ``radial'') Baryon Acoustic Oscillations
(BAO), the differential ages of galaxies (DAG), as well as the redshift drift of the spectral lines (RDSL) will constrain $z_t$,  thereby helping
to narrow the parameter space for the most realistic models describing the accelerating Universe.

The article is structured as follows. In Sect. II, we discuss the general problem of the transition redshift for different cosmologies. In Sect. III,  we discuss how it can be thought as a new cosmic parameter even in the context of the $\Lambda$CDM model. In section IV, the transition redshit is discussed as a new cosmological number in the sense of Sandage. The   possibility to access it from independent observations  is also discussed in the corresponding subsections. Finally, in the conclusion section we summarize the basic results.

\begin{figure}
\label{fig1.eps}
\vspace{-.15in}
\centerline{\epsfig{figure=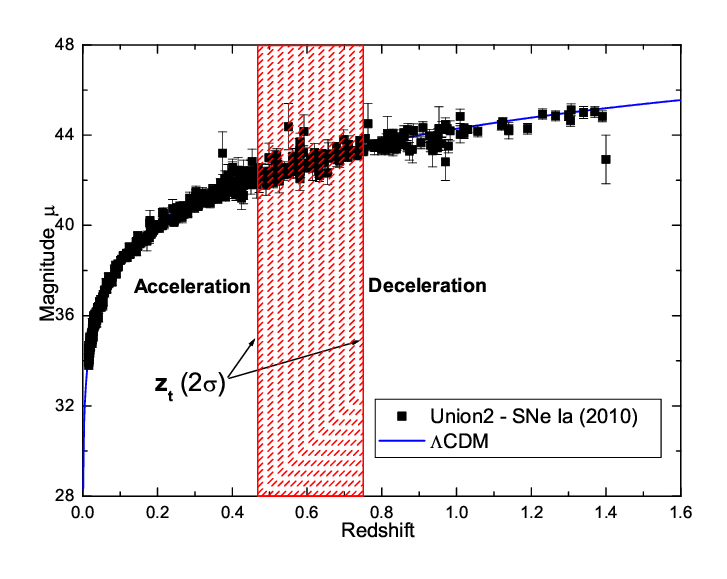,width=3.4truein,height=2.8truein}
\hskip 0.05in}
\caption{The relative magnitude as a function of the redshift for the SNe Ia sample compiled by Amanullah et al. \cite{Union2}. The vertical strip shows the Riess {\it et al.} \cite{Riess07} limits on the transition redshift, $z_t=0.426^{+0.27}_{-0.089}$ (at 95\% c.l.), based on a kinematic approach \cite{TurRies02,CunhaLimaMNRAS08,GCL09,Lu2011}.
It will be argued here that the transition redshift will become accessible by future observations, and, as such, it may play the role of a primary cosmological parameter.}
\end{figure}      
\section{Transition redshift in FRW Geometries}

In what follows we restrict our attention to the
class of spacetimes described by the FRW line element (unless explicitly stated we set $c=1$):
\begin{equation}
ds^2 = dt^2 - a^{2}(t) \left[\frac{dr^{2}}{1-kr^2} + r^{2}(d \theta^2 + \rm{sin}^{2} \theta d \phi^{2})\right],  
\end{equation}
where $a(t)$ is the scale factor and $k$ is the curvature constant,
which can be $-1$, $0$, or $1$, for a spatially open, flat or closed
Universe, respectively. Although inflationary models and recent
observations from cosmic microwave background (CMB) temperature anisotropies 
favor a spatially flat Universe, we shall not
restrict ourselves to this case.

In this background, the Einstein Field Equations (EFE) and the decelerating parameter, $q$, can be written as:

\begin{equation}
\left(\frac{\dot{a}}{a}\right)^2 + \frac{k}{a^2}=\frac{8\pi G}{3}\rho_T,
\end{equation}
\begin{equation}
\frac{\ddot{a}}{a}=-\frac{4\pi G}{3}(\rho_T+3p_T),
\label{accel}
\end{equation}
\begin{equation}
q (z) \equiv -\frac{a\ddot{a}}{{{\dot a}^{2}}}= -\frac{1}{H^{2}}\left(\frac{\ddot{a}}{a}\right),
\end{equation}
where $\rho_T$ and $p_T$ are the total energy density and pressure of the mixture and $H(t)={\dot a}/a$ is the Hubble parameter.
Note that the acceleration equation (\ref{accel})
does not depend explicitly on the curvature, and, similarly, the same
happens with the transition redshift ($z_t$) since it is implicitly defined by
the condition $q(z_t)=\ddot{a}(z_t)=0$. It is also worth noticing that all kinematic approaches
developed in the literature \cite{TurRies02,CunhaLimaMNRAS08,GCL09,Lu2011} point to
a transition redshift in the past, that is, at intermediate redshifts ($z_t < 1$). The importance of such results comes from the fact that the derived transition redshift is independent
of any dark energy models as well as of the underlying gravity theory.
However, as remarked by Catt{o}en and Visser \cite{CW2007}, some caution  is needed when the expansions are combined with high-$z$ data in the framework of  kinematic approach.  In order to cure this problem they also suggested an extended $y$-redshift expansion which is able to fix mathematically the correct convergence radius. Naturally, the price for avoiding convergence issues for the expansion in
powers of $z$ has severe consequences on model predictability, but this could be overcome by
a much larger SN Ia sample (which is foreseen in future surveys), by the use of other kinds of
data and priors, and maybe by a smarter (optimal) cosmographic modeling. More recently,  the $y$-redshift expansion was adopted by 
Guimar\~aes and Lima \cite{GL11} in order to discuss the possibility of a decelerating stage  in the future of the Universe. 

As widely known, the first SNe Ia analyses were done assuming a
constant $\Lambda$ for the dark energy component. However,  due to the coincidence and cosmological constant problems
several candidates for dark energy were proposed in the literature. At
present, beyond the cosmological constant there is a plethora of relativistic
dark energy candidates capable to  explain the late time accelerating
stage, and, as such,  the space parameter of the basic observational
quantities is rather degenerate. The most economical explanation
is provided by the flat $\Lambda$CDM model which has only one dynamic free
parameter, namely, the vacuum energy density. It seems to be consistent with
all the available observations provided that the vacuum energy density is fine
tuned to fit the data ($\Omega_{\Lambda} \sim 0.7$).  However, even considering that the
addition of extra fields explain the late time accelerating stage and
other complementary observations \cite{CMB,Clusters,BAO,BAO1}, the need of
(yet to be observed) dark energy component with unusual properties is
certainly a severe hindrance.

For the sake of simplicity, next section  we
focus our attention on the $\Lambda$CDM model and its
predicted transition redshift. The main aim is to show how to built a complementary space parameter based on
the transition redshift as a basic quantity. Further, it will be discussed how such an approach  may  be useful
to discriminate the realistic accelerating world models proposed in the literature.

\section{Transition Redshift in $\Lambda$CDM Models}

The late time observed Universe in $\Lambda$CDM models is composed almost completely of
pressureless matter (consisting of dark and normal baryonic matter
components), and a negative pressure cosmological constant energy
density, since the radiation contribution at low redshifts is just
$\sim 10^{-5}$ of the total energy density. Following standard lines, we write deceleration parameter as

\begin{equation}
q (z) \equiv
-\frac{1}{H^{2}}\left(\frac{\ddot{a}}{a}\right)=\frac{(1+z)}{H(z)}\frac{dH(z)}{dz} - 1, \label{e2}
\end{equation}
with the Hubble parameter assuming the form below
\begin{equation}
H(z) = H_0\left[\Omega_M (1 + z)^{3} + \Omega_{\Lambda} + \Omega_k (1 + z)^{2})\right]^{1/2},
\label{Heq2}
\end{equation}
where $\Omega_M$, $\Omega_{\Lambda}$ and $\Omega_k = 1 -\Omega_m - \Omega_{\Lambda}$, are the present day matter, vacuum and curvature parameters,  respectively.

In this framework, the  acceleration equation (3) yields
\begin{equation}
\label{AccelLcdmZ}
\frac{\ddot{a}}{a}=-\frac{4\pi G}{3}\left[\rho_{M}(1+z)^3-2\rho_\Lambda\right],
\end{equation}
while  the transition redshift (where $\ddot{a}$ vanishes)  can be written as
\begin{equation}
\label{ZtLcdm}
z_t=\left[ {\frac{2\rho_\Lambda}{\rho_{M}}} \right]^\frac{1}{3}-1=\left[ {\frac{2\Omega_{\Lambda}}{\Omega_{M}}}\right]^\frac{1}{3} -1.
\end{equation}

\begin{figure}
\label{fig2.eps}
\centerline{\epsfig{figure=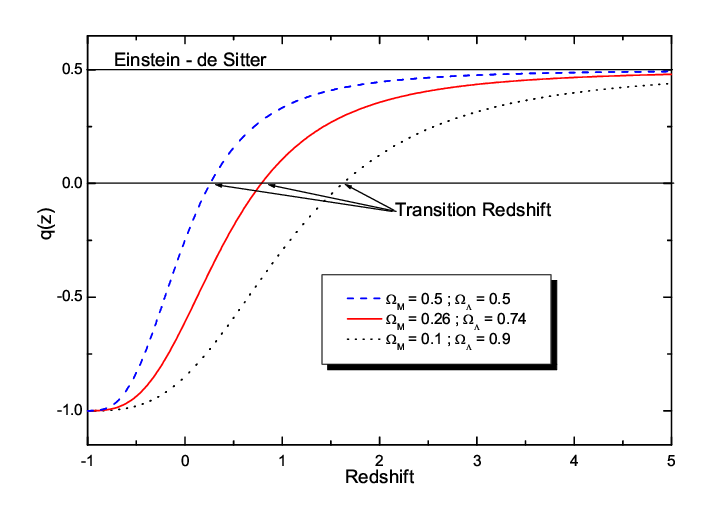,width=3.4truein,height=2.8truein}
\hskip 0.05in}
\caption{Deceleration parameter as a function of the redshift for a flat $\Lambda$CDM model and some selected values of $\Omega_M$. The solid (red) curve is the evolution of $q(z)$ for the so-called cosmic concordance model. The transition redshift is heavily dependent on the possible values of the density parameter, and, as expected, $z_t$ is higher for smaller values of $\Omega_m$.}
\end{figure}
As should be expected,  the transition redshift does
not depend explicitly on the curvature, only on the ratio of vacuum
density and matter density. However, if we assume a spatially flat
Universe, we obtain the normalization condition
$\Omega_M+\Omega_\Lambda=1$ and Eq. (\ref{ZtLcdm}) becomes
\begin{equation}\label{flatzt}
z_t= \left[{\frac{2(1-\Omega_M)}{\Omega_M}}\right]^\frac{1}{3} -1.
\end{equation}

In Figure 2, we show the deceleration parameter of a spatially flat $\Lambda$CDM model as a function of the redshift for some
selected values of the matter density parameter. Note that the possible values of $z_t$ are strongly dependent on the values assumed for $\Omega_M$.

\begin{figure}
\label{WmZtLCDM}
\centerline{\epsfig{figure=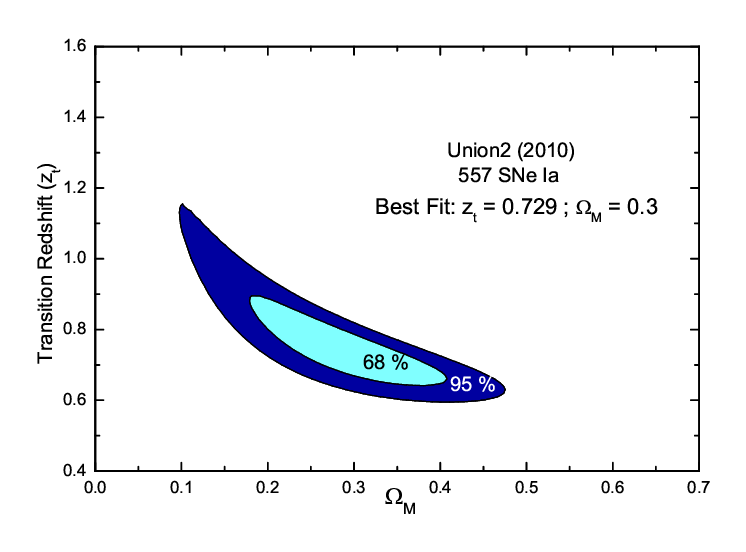,width=3.4truein,height=2.8truein}
\hskip 0.05in}
\caption{Transition redshit as a function of $\Omega_M$ for a general $\Lambda$CDM model ($\Omega_k \neq 0$).  The inner and outer contours (68\%, 95\%) C. L. show the limits on the transition redshift based on the  SNe Ia sample compiled by Amanullah {\it et al.} \cite{Union2}. The best fit values are explicitly shown in the plot.}
\end{figure}

But how can we built a suitable space parameter having the transition redshift as a basic variable? In principle, even before to discuss the related 
observations, we observe that such a bidimensional parameter space, say,  ($\Omega_M, z_t$) can be defined through the transition redshift itself. In the 
framework of a general $\Lambda$CDM model, one may combine  Eqs. (\ref{ZtLcdm}) and (\ref{Heq2}) in order to obtain an expression for $H(z_t,\Omega_M)$.

In Figure 3, we display our $\chi^{2}$-statistical analysis in the plane ($\Omega_M, z_t$) based on the supernova sample (Union2) as compiled by Amanullah 
and collaborators \cite{Union2}. Note that for a general $\Lambda$CDM model, the transition redshift is well constrained at $2\sigma$ confidence level. More 
precisely, we have found a transition redshift on the interval  $0.60 \leq z_t \leq 1.18$ (2$\sigma$, joint analysis).  Indirectly, this result shows that the cosmic 
expansion history can also be rediscussed in terms of the transition redshift.    

\begin{figure}
\vspace{.15in}
\centerline{\epsfig{figure=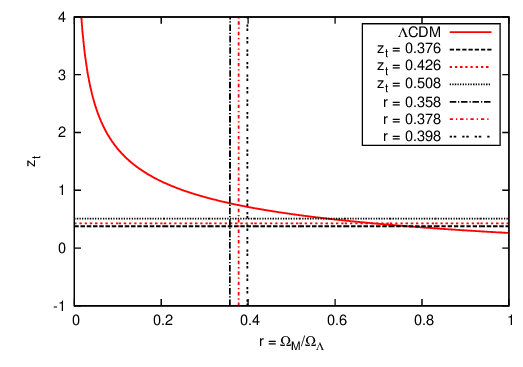,width=3.5truein,height=2.8truein}
\hskip 0.05in}
\caption{\label{FigZtLcdmNonflat} Transition redshift as a function of the density ratio $r=\Omega_M/\Omega_\Lambda$. Also shown are the Riess et al. 
\cite{Riess07} 68\% c.l. limits on the transition redshift, $z_t=0.426^{+0.082}_{-0.050}$ and the derived WMAP5 68\% c.l. on the ratio, $r=0.387\pm0.020$.}
\end{figure}

In Figure 4, we have plotted the dependence of \zt\ as a function of the ratio, $r=\Omega_M/
\Omega_\Lambda$, for a general $\Lambda$CDM model. The horizontal and vertical strips correspond, respectively,  to $z_t=0.426^{+0.27}_{-0.089}$ (2$\sigma$) from Riess et al. \cite{Riess07}  and the derived WMAP5 68\% c.l. on the ratio, $r=0.387\pm0.020$.

\begin{figure}
\vspace{.15in}
\centerline{\epsfig{figure=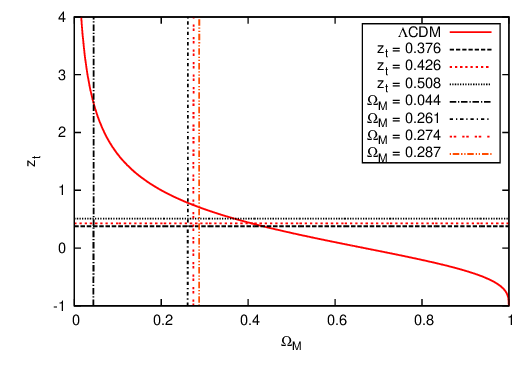,width=3.5truein,height=2.8truein}
\hskip 0.05in}
\caption{\label{FigZtLcdmFlat} Transition redshift as a function of the matter density parameter for a spatially flat Universe. Also shown are the Riess et 
al. \cite{Riess07} 68\% c.l. limits on the transition redshift, $z_t=0.426^{+0.082}_{-0.050}$ and the derived WMAP7 68\% c.l. on the matter density 
parameter, $\Omega_M=0.274\pm0.013$.}
\end{figure}

In Figure (\ref{FigZtLcdmFlat}) we display the transition redshift for the flat case as a function of the matter density parameter. As expected, in the limit $\Omega_M \rightarrow 1$ (Einstein-de Sitter model) there is no transition. The horizontal lines in both plots are the kinematic limits  on $z_t$ derived by Riess et al. \cite{Riess07}, $z_t=0.426^{+0.27}_{-0.089}$ ($2\sigma$), by using a linear parametrization of the deceleration parameter
$q(z)$ \cite{TurRies02}.   More recently, one of us have checked their analysis \cite{CunhaLimaMNRAS08} and have found
$z_t=0.426^{+0.082+0.27}_{-0.050-0.089}$, at 68\% and 95\% c.l.,
respectively, consistent with their result. For other parameterized deceleration parameter appeared in Ref. \cite{Lu2011}, the transition redshift is
constrained with $z_t=0.69^{+0.23}_{-0.12}$  and $z_t=0.69^{+0.20}_{-0.12}$. The interest of such an approach is that it holds regardless of the
gravity theory.
The vertical lines represent the  constraints derived by the WMAP7 team \cite{WMAP7} through a joint analysis involving
CMB, BAO and $H_0$ (Table 14, RECFAST version 1.5 \cite{WMAP7}).  The limits at 1$\sigma$ c.l. for  the density parameter and density ratio are
$\Omega_M=0.274\pm0.013$, $r\equiv\Omega_M/\Omega_\Lambda=0.387\pm0.020$, respectively.

As one may see from these figures, the standard concordance flat
$\Lambda$CDM model is just marginally consistent with the transition
redshift derived from the kinematic approach of Riess et al. \cite{Riess07}. A
fortiori, this could be seen to raise some mild flags with the
standard $\Lambda$CDM model, to add to the more well-known
cosmological constant problem (CCP) and coincidence problem. Nevertheless,  we observe that the recent CMB results released by Planck Collaboration \cite{Planck} provided a best fit value of  $\Omega_M$ higher than the previous one found by WMAP thereby diminishing the value of the transition redshift. In the flat case, for instance, it was found $\Omega_M = 0.314\pm 0.020$ and from Eq. ({\ref{flatzt}}), the corresponding transition redshift goes to $z_t=0.632\pm 0.043$ and, as such, slightly displaced towards the interval predicted by some kinematic analyzes.  

Thus, we can see the importance of the transition redshift in order to
distinguish among similar dark energy models and as consistency check
for any new dark energy model. One could rule out, for example,
cosmological models with no transition redshift at all, as the family of dust filled
FRW type models, and some subclasses of $\Lambda(t)$CDM models \cite{decaying1},
or more generally some coupled dark matter-dark energy models.  

\section{Transition Redshift as a New Cosmic Discriminator}

In the early seventies, Alan Sandage \cite{Sandage70} defined Cosmology
as the search for two numbers: $H_0$ and $q_0$. In the conclusions of the paper,  
by commenting about future observational values of $H_0$ and $q_0$, he wrote: ``The present discussion is only a
prelude to the coming decade. If
work now in progress is successful,
better values for both $H_0$ and $q_0$ (and
perhaps even $\Lambda$) should be found, and
the 30-year dream of choosing between
world models on the basis of
kinematics alone might possibly be
realized''.  Indeed, it was needed to wait for almost 3 decades to obtain such quantities with great precision by using  SNe Ia
as standard candles \cite{const}. Fortunately, Sandage lived enough to see his predictions substantiated by the new observational techniques. 

Now, based on the results about the transition redshift presented in the earlier sections (see Figs. 1 and 2),  
let us discuss (from a more observational viewpoint) the possibility to enlarge Sandage's vision by including  the
transition redshift, $z_t$, as the third cosmological number.
To begin with, let us observe that the general expression for $q(z)$ as given by Eq. ({\ref{e2}) means that the transition redshift can empirically be defined as:

\begin{equation}\label{z_tM}
z_t = \left[\frac{d {\ln} H(z)}{dz}\right]^{-1}_{|_{z=z_t}} - 1.
\end{equation}
Therefore, from a formal viewpoint we may access the value of $z_t$ through  a determination of $H(z)$ at least around a redshift interval involving the transition redshift. 

How can this be worked out? To begin with we observe that the determination of $z_t$ is fully equivalent to the reconstruction of $q(z)$  from  $H(z)$ data and its first derivative (see Eq. (\ref{e2})). If we have enough $H(z)$ data, such a reconstruction can be improved, as shown by Carvalho and Alcaniz \cite{CarvalhoAlcaniz11} based on Monte Carlo  simulations. In fact, the $H(z)$ data sample just keeps increasing, as it has been found already 34 Hubble parameter values on the literature \cite{Sharov14}. Rigorously, one needs only data involving the interval where $q(z)$ changes its sign. As we know, this occur at redshift of the order of unity  and the $H(z)$ data sample already has many values on such an interval \cite{FarooqRatra13,Sharov14}.

In what follows, we suggest that at least 3 different kinds of ongoing and future observations  can be used to obtain $H(z)$, and, therefore, the transition redshift, namely: (i) the line-of-sight (or ``radial'') baryon  acoustic oscillations (BAO), (ii) the differential ages of galaxies (DAG), and
(iii) the redshift drift of the spectral lines (RDSL). As we shall see, potentially, all these techniques (together or separately)  are able to provide the
value of $z_t$ from the related $H(z)$ measurements. Still more important, the accuracy on the measurements of $H(z)$ must increase thereby allowing the observers to determine a value of $z_t$ that could be useful as a robust cosmic discriminator in the near future.    

\subsection{Radial BAO}

BAO in the last scattering surface provide statistical standard rulers in the late time cosmic structures of known physical lengths
thereby making such measurements important cosmic probes. The first measurement of the BAO acoustic peak was obtained by Eisenstein et al. \cite{BAO} through a spherical averaged two-point correlation function from luminous red galaxies data compiled by the Sloan Digital Sky Survey (SDSS).
This first measurement was an average between the so called transversal (or angular) BAO, $\sigma_{BAO}$, measured in the plane of the sky, and the radial BAO, $\pi_{BAO}$, which is measured along the line of sight. Their importance as new standard rulers have also been confirmed by many independent studies \cite{BAO1,Gast2009,Basset}. Although still controversial \cite{Kazin2010}, some authors have argued that the redshift space distortions might boost the radial BAO signal \cite{Thian2011,JPAS}.
However,  the transversal BAO which provides a direct information on the angular distance is not particularly useful to determine the $H(z)$ function, and, consequently, the transition redshift ($z_t$).

On the other hand, the line-of-sight BAO yields an indirect measurement of $H(z)$ because it is related with the expansion history by the expression:  
\begin{equation}
\pi_{BAO}=\frac{c\Delta z}{H(z)},
\end{equation}
and, hence, the values of $H(z)$ can be inferred because $\pi_{BAO}$ and $\Delta z$ are observationally determined.  As a matter of fact, determinations  of
 the radial BAO  have already been used to extract some $H(z)$ values at low redshift \cite{Gast2009}. However, the situation can even be improved with the
 planned operation of instruments dedicated to BAO measurements. In principle, a BAO survey require redshift accuracy and coverage of enough volume (and
 area). In this way,  an ideal instrument is needed to have a large mirror size allied to a capability for producing simultaneously a large number of spectra,
 for instance, through a Wide Field Multi-Object Spectrograph.  There are several possibilities somewhat related with the late stages of the DFTE report
 (for a detailed review see Basset and Hlozek \cite{Basset}).

An alternative  possibility is the so-called PAU survey \cite{Benitez10} (now called JPAS \cite{JPAS}) based on photometric instead of spectroscopic redshifts. The basic 
idea is that  photometric redshifts of galaxies with enough precision to measure BAO along the line of sight may become available even with a 2.5m telescope.  Their proponents claim that the survey will produce a unique data set in the optical wavelength for all objects in the north sky up to $m_B=23-23.5$ arcsec$^{-2}$ ($5\sigma$) thereby making the JPAS very competitive in comparison with other (photometric or spectroscopic) ground-based BAO surveys \cite{Benitez10}.  

Summing up, one may expect that  radial BAO measurements from luminous red galaxies and other objects with different techniques will be able to provide the instantaneous expansion rate $H(z)$ at intermediate redshifts, and, as argued here (see Eq. ({\ref{z_tM})),  the transition redshift itself.

\subsection{Differential Ages of Galaxies (DAG)}
In  terms of the redshift, it is easy to show that the Hubble parameter, $H(z)$, can be expressed as:
\begin{equation} \label{state}
H(z) = -\frac{1}{(1+z)} \frac{dz}{dt}.
\end{equation}
This simple expression means that  measurements of the differential redshift ages for a
class of objects, $dz/dt$, potentially, provide a direct estimate of $H(z)$, and, therefore, is also an interesting  observational window to access the transition redshift.  
As recently discussed by several authors, age differences between
two passively evolving
galaxies formed at the same time but separated
by a small redshift interval have already been inferred.
In principle, the statistical significance
of the measurement can be improved by selecting fair samples
of passively evolving galaxies at the corresponding redshifts, and
by comparing the upper cutoff in their age distributions.

In a point of fact, by choosing carefully a sample of old elliptical galaxies (similar metallicities and low
star formation rates), Jimenez and collaborators used this method to obtain the first determination
of the curve $H(z)$ \cite{Jimenez2003}. At present, only eleven $H(z)$ data have been inferred with such a technique, some of them at high redshifts (up to $z = 1.75$).
In this line,  Stern et al. \cite{Stern10} also compiled
an expanded set of $H(z)$ data (see their Figure 13) and
combined it with CMB data in order to constrain dark energy parameters
and the spatial curvature. The amount of $H(z)$ data keep increasing, as more Hubble parameter values have been added to the list \cite{FarooqRatra13,Sharov14}. More recently, simulations using Monte Carlo Technique based
on $\Lambda$CDM model have also been discussed by several authors in order to constrain cosmological parameters \cite{MaZhang,CarvalhoAlcaniz11}.  
Therefore, similarly to what happens with BAO measurements, it will be  possible in the near future to obtain the expansion rate history $H(z)$ from DAG with
percent precision that will translate into measurement of $z_t$ with few times the same precision.

\subsection{Redshift Drift}

Some decades ago, Sandage ~\cite{san} proposed that the dynamical expansion history  could  directly
be traced by the time evolution of redshift (now usually called redshift drift, ${\dot z}$).  The expansion of the
Universe is expressed by the scale factor, $a(t (z)) = (1 +z)^{-1}$. Therefore,
the time evolution of the scale factor, or change in redshift,
${\dot z}$,  directly measures the expansion rate of the Universe. 
In other words, neglecting effects of peculiar velocities, the redshift of a comoving  object is a function of the observing time, and, us such, its value in the future will be different of what is measured today. As shown long ago by McVittie \cite{mc}, it can be expressed as  

\begin{equation}
\dot{z }(z) = \frac{d z}{dt_{{o}}} (t_{o}) = (1 + z )H_0  - H(z),
\end{equation}
where $t_{o}$ denotes the present day observing time.  
This clearly shows that measurements of ${\dot z}$  are able to 
provide the Hubble parameter at redshift $z$. This redshift drift 
signal, ${\dot z}$, is indeed  very small and barely accessible from techniques available at Sandage's time.  
It should be stressed that the redshift drift is a direct 
and fully model independent measurement of $H(z)$ since apart the FRW metric 
it does not require any cosmological assumption.  The significance of this tool has
been recently rediscussed by several authors by taking into 
account the current (and near future) observational capabilities ~\cite{ot,lo,joe,jain2010}.
By defining the apparent velocity shift, ${\dot {\mathrm V}}= \Delta {\mathrm v}/\Delta t_0$, 
it is easy to check that the redshift drift can also be 
rewritten as (see, for instance, Refs. \cite{joe,jain2010})
\begin{equation}
\frac{\dot {\mathrm V}(z)}{c} = \frac{(1+z)H_0
-{H(z)}}{1+z}.
\end{equation}
In order to use properly the above results, the first task is to identify  the classes  of objects and the corresponding spectral properties suitable for ${\dot V}$ measurements.  By assuming  high-z observations ($z \sim 4$) for a decade, Loeb \cite{lo} estimated  $\Delta{\mathrm v} \approx 6 $ cm/s in the framework of a  $\Lambda$CDM model. He claimed  that such a weak signal would be measured by using the absorption lines (Ly$\alpha$ forest) observed in the spectra of quasi-stellar objects (QSOs). 

More recently,  Linske et al. \cite{joe} argued that Ly$\alpha$ in the redshift range $(2 \leq z \leq 5)$ are indeed the most convenient targets.  In their very detailed study, the possibility of Ly$\beta$ forest and the influence of peculiar motions were also investigated.  They also discussed the possibility  of detecting and characterizing the cosmological redshift drift based on the next generation of extreme large telescope (ELT). In particular, a velocity drift experiment over 20 years using 4000 hours of observing time on a 42 meters ELT would be able to exclude $\Omega_{\Lambda}=0$ with 98.1 per cent confidence level.  It should be noticed that redshift drift measurements  constrain $H(z)$ at high-z and, as such,  would complement current and future data based on  SNe Ia, radial BAO, and differential ages of galaxies.

Finally, we also observe that the idea of a direct accelerating probe (or the transition redshift) has also been discussed by Sahni, Shafieloo and Starobinsky \cite{Sahni} based on the expression of the average decelerating parameter $\bar q$ defined by Lima \cite{Lima} in connection with the total age of the Universe. By calculating the value of $\bar q$ over a small redshift range close to $z_t$ they argued  that measurements of the Hubble parameter used to determine the cosmological redshift at which the universe began to accelerate, without reference to the matter density parameter. As it appears, the concept of the acceleration probe involves only the reconstruction of $H(z)$, and, therefore, it is also useful to find $z_t$ in a model independent manner because it does not need higher derivatives of the data.

\section{Conclusion}

In this paper, we have advocated that the transition redshift should be considered as a primary cosmic parameter in the sense of Sandage \cite{Sandage70}. By starting with a generic $\Lambda$CDM cosmology, it was shown  how a convenient parameter space ($\Omega_M, z_t$), involving directly the transition redshift, could be built. As it appears, such an analysis might  be extended for other relativistic models or even for accelerating cosmologies based on modified gravity theories. It was also argued that the transition redshift will be directly accessed trough measurements of $H(z)$ by the ongoing and future observational projects. In particular, we have discussed the most promising ones involving the line-of-sight (or ``radial'') baryon acoustic oscillations,  the differential age of galaxies, as well as the redshift drift of the spectral
lines. Potentially, the work now in progress allied with the near future observations may transform  the transition redshift in an interesting primary cosmic variable.

\vspace{0.05cm}
\vspace{0.1cm}{\bf Acknowledgments:} The authors are grateful to Varun Sahni for helpful correspondence. J.A.S.L. is partially supported by CNPq and FAPESP, by the time of writing this paper, J.F.J. was supported by FAPESP grant 2010/05416-2 by the time of writing this paper, J.F.J. and RCS are grateful to INCT-Astrof{\'i}sica and the Departamento de Astronomia (IAG-USP) for hospitality and facilities.


\end{document}